\documentclass[10pt,article,superscriptaddress,aps,twocolumn,pra,longbibliography]{revtex4-2}

\usepackage{amssymb,amsmath,amsbsy,amsgen,amsfonts,amsthm,mathtools,mathrsfs,amstext,bm}
\usepackage{array}
\usepackage{color}
\usepackage{hyperref}
\usepackage{epstopdf,graphicx}
\usepackage{verbatim}
\usepackage{url}
\usepackage{booktabs}
\usepackage{quantikz}
\usepackage{capt-of}
\allowdisplaybreaks[1]

\begin{document}

\title{Quantum circuit optimization using deep reinforcement learning: Applications across multiple gate sets}

\author{Khoa Dang Tao}
\affiliation{Department of Physics, Hanyang University, Seoul 04763, Republic of Korea}

\author{Sumin Jin}
\affiliation{Department of Physics, Hanyang University, Seoul 04763, Republic of Korea}

\author{Muhammad Raza}
\affiliation{Department of Physics, Hanyang University, Seoul 04763, Republic of Korea}

\author{Changhyoup Lee}
\email{changhyoup.lee@gmail.com}
\affiliation{Department of Physics, Hanyang University, Seoul 04763, Korea}
\affiliation{Hanyang Institute for Quantum Science and Quantum Technology, Hanyang University, Seoul 04763, Korea}
\affiliation{Research Institute for Natural Sciences, Hanyang University, Seoul 04763, Korea}
\date{\today}

\begin{abstract}
The practical implementation of quantum algorithms on noisy intermediate-scale quantum devices encounters operational limitations due to decoherence and other sources of noise inherent in real hardware. To mitigate these errors while preserving the original functionality of the algorithm, shorter quantum circuits are therefore preferred. This motivates the development of effective quantum circuit optimization algorithms. Learning-based approaches have emerged as a leading candidate, yet existing autonomous agents remain inefficient, spending most of their training capacity rediscovering elementary reductions that deterministic rule-based methods already handle reliably.
To address this challenge, we propose a reinforcement learning framework that embeds a deterministic Commutation-and-Reduction (CR) algorithm directly into the training environment. After every agent action, the CR algorithm automatically resolves elementary commutations and cancellations, enabling the agent to focus its learning capacity on the non-trivial optimizations where reinforcement learning adds real value. Empirical evaluation across two gate sets, the universal Clifford+T basis and the CNOT+Pauli basis, shows that RL+CR produces shorter circuits than a standard RL agent at all tested scales. We demonstrate that RL trained on smaller quantum circuits can be applied to larger quantum circuits. On 20-qubit Clifford+T circuits, five times larger than the training circuits, RL+CR removes twice as many gates as standard RL. This work provides a robust approach that could accelerate the compilation and optimization processes for future fault-tolerant and utility-scale quantum systems. 
\end{abstract}

\maketitle

\section{Introduction}\label{sec_intro}
Quantum computing has emerged as a transformative computational paradigm, promising to solve problems that are fundamentally intractable on classical hardware~\cite{bauer2020quantum, montanaro2016quantum, dalzell2023quantum, arute2019quantum, kim2023evidence}. Realizing this promise on near-term hardware, however, remains constrained by the physical limitations of noisy intermediate-scale quantum devices, where decoherence, gate infidelities, and restricted qubit connectivity collectively degrade the reliability of quantum algorithms~\cite{preskill2018quantum, kjaergaard2020superconducting, linke2017experimental, bharti2022noisy}. Because both the depth and the gate count of a quantum circuit directly amplify these errors, the executable complexity of any algorithm is strictly bounded by the structural cost of its circuit representation~\cite{murali2019noise}. While quantum error correction offers a principled long-term solution, its substantial qubit and gate overhead place it beyond the reach of current hardware~\cite{fowler2012surface, google2023suppressing, google2025quantum}. Quantum circuit optimization (QCO), which compresses a circuit while preserving its underlying unitary, has therefore emerged as a critical step in bridging the gap between abstract algorithms and physical execution.

Existing QCO frameworks fall into two dominant paradigms. Rule-based compilers apply local template matching and gate commutation to achieve fast, reliable reductions~\cite{sivarajah2021t, aleksandrowicz2019qiskit, nam2018automated}. However, their reliance on static, hand-designed heuristics limits their ability to generalize across diverse circuit architectures. Algebraic methods built on the ZX-calculus~\cite{duncan2020graph} enable global graph-level simplifications, yet the subsequent extraction of a hardware-compatible circuit often reintroduces complexity that negates the gains achieved in the graph domain~\cite{backens2021there}.

Reinforcement learning (RL) offers a natural alternative by treating circuit optimization as a sequential decision-making problem in which an agent learns, through experience~\cite{fosel2021quantum, moro2021quantum, ostaszewski2021reinforcement, kremer2024practical}. Despite its conceptual appeal, a direct RL formulation suffers from a fundamental inefficiency. Most of the agent's training capacity is consumed rediscovering elementary reductions that deterministic rule-based methods already handle reliably, while the non-trivial, multi-step optimizations where RL would genuinely add value remain out of reach.

Here, we propose a reinforcement learning framework that resolves this inefficiency by embedding the Commutation-and-Reduction (CR) algorithm directly into the training environment. After every agent action, the CR algorithm exhaustively performs the trivial commutations and cancellations that a rule-based optimizer would immediately identify, returning only the simplified circuit to the agent. The agent therefore never observes a circuit with elementary simplifications, thus never expends learning capacity rediscovering them. Its entire learning effort is concentrated on the non-trivial transformation strategies.

\begin{figure*}[t]
    \centering
    \includegraphics[width=0.8\linewidth]{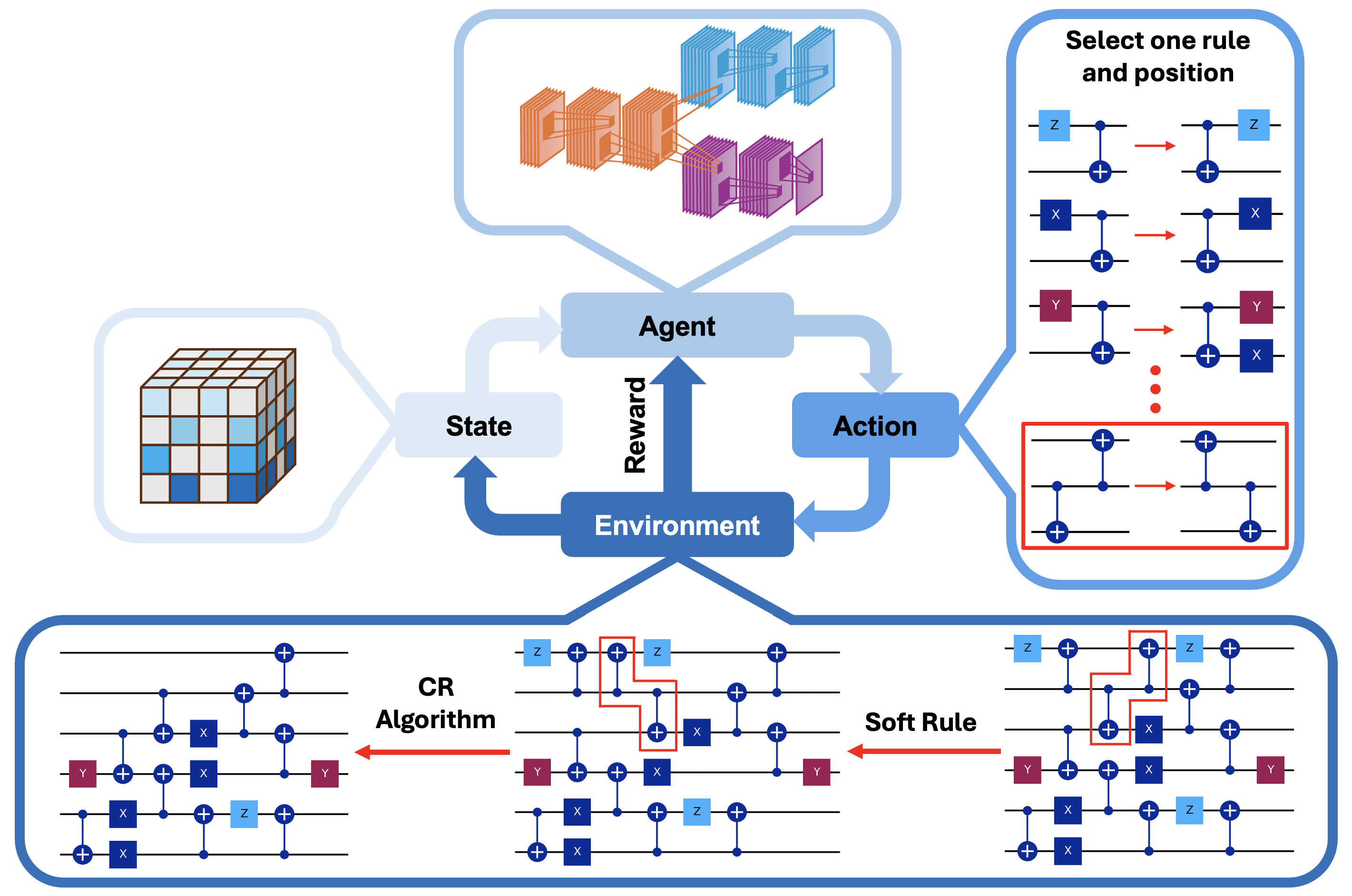}
    \caption {Overview of the integrated reinforcement learning framework. The diagram illustrates the closed-loop interaction between the Actor–Critic agent and the environment.}
    \label{fig:framework}
\end{figure*}

We evaluate this framework on two structurally distinct gate sets: the universal Clifford+T basis~\cite{amy2014polynomial, kissinger2019reducing}, and the CNOT–Pauli basis, hereafter referred to as CXP. Across all tested scales, the proposed method produces shorter circuits than standard RL baselines. This work establishes a robust approach to circuit optimization that could accelerate the compilation pipelines of fault-tolerant and utility-scale quantum systems.

\section{Method}\label{sec_method}

We formulate quantum circuit optimization as a Markov decision process~\cite{sutton1998introduction} and solve it with a deep reinforcement learning agent trained by Proximal Policy Optimization (PPO)~\cite{schulman2017proximal}. The central methodological contribution is the integration of a CR algorithm inside the environment’s transition function, which removes elementary reductions before the agent observes the next state. This restricts the agent’s learning capacity to non-trivial transformations that lie beyond the reach of greedy rule-based heuristics.

\subsection{Framework Overview}
The complete optimization framework is summarized in Figure \ref{fig:framework}. The agent and the environment interact in a closed loop, with the quantum circuit itself acting as the state that is iteratively refined. At each step $t$, the environment encodes the current circuit as a three-dimensional tensor $s_t$. The Actor head of the agent processes $s_t$ through a fully convolutional Actor–Critic network and outputs a joint distribution over the action triple $a=(q, m, k)$, where $q$ is the target qubit, and $m$ is the target moment, and $k$ indexes a transformation rule in a set of soft and decomposition rules $\mathcal{R}$. After masking out impermissible actions, the agent samples a single triple, and the environment applies the corresponding rule to produce an intermediate circuit. This intermediate circuit is then passed to the CR algorithm, which deterministically performs all commutations and hard-rule cancellations that the chosen rule has exposed. The output of CR defines the next state $s_{t+1}$, and the reward $r_{t+1}$ is computed from the difference in circuit cost between $s_t$ and $s_{t+1}$. Trajectories are aggregated across multiple environments and used to update the network with PPO. The remainder of this section describes each component of the framework in turn: the transformation rules that constitute the action space, the CR algorithm that resolves trivial reductions deterministically, the state representation, the network architecture, the reward, and the training pipeline.

\begin{figure*}[t]
    \centering
    \includegraphics[width=1.0\linewidth]{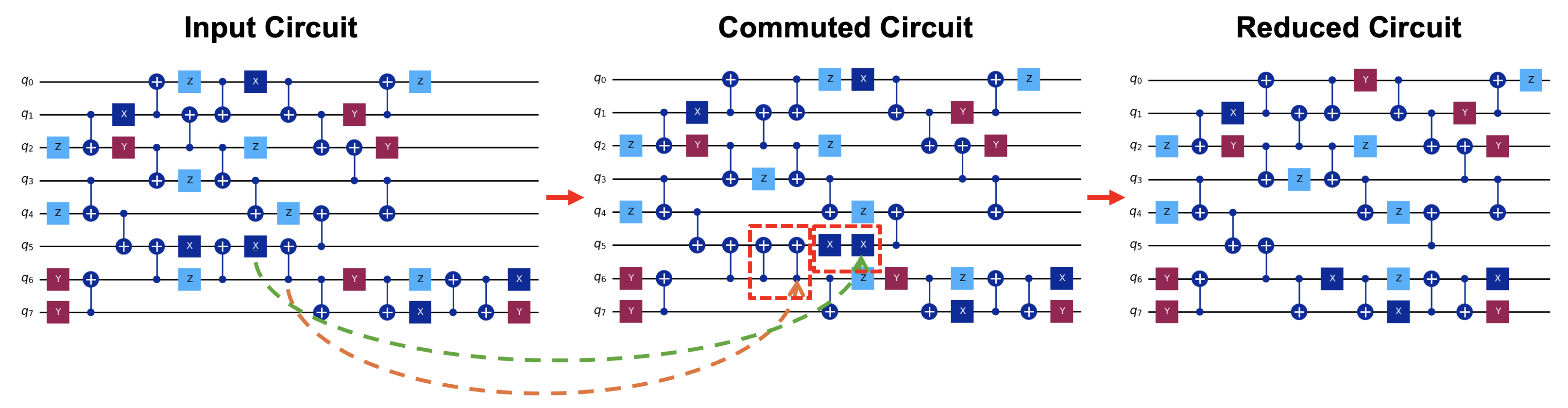}
    \caption {The two-phase execution of the Commutation-and-Reduction algorithm. The input circuit contains a disordered distribution of gates. Dashed arrows trace representative gates during the commutation phase: X, Y, and Z gates propagate rightward, while a CNOT propagates leftward. The exposed X-X and CNOT-CNOT pairs (red box) cancel to the identity (I) in the reduction phase, producing the reduced circuit.}
    \label{fig:CR}
\end{figure*}

\subsection{Quantum Circuit Transformations}

Quantum circuit optimization proceeds by applying local rewrites that preserve the implemented unitary while altering the structure of the circuit. Three classes of equivalence-preserving transformations are used. \textit{Hard rules} strictly reduce the gate count or depth and are unconditionally beneficial, such as self-inverse cancellation ($OO^\dagger = I$). \textit{Soft rules} preserve gate count but exploit commutation relations $[A, B] = 0$ to reorder gates and expose hidden cancellations. \textit{Decomposition rules} temporarily increase the local gate count by rewriting a pattern as an equivalent longer sequence, in order to expose structural symmetries that subsequent soft and hard rules can exploit. The complete rule sets for both bases are listed in Appendix \ref{app:rules}.

Because hard rules are unconditionally beneficial, applying them never requires strategic judgement. They are therefore excluded from the agent’s action space and delegated to the environment. The action space consists of soft and decomposition rules, whose value depends on context.

\subsection{Commutation-and-Reduction Algorithm}
The CR algorithm is a deterministic routine that exhaustively applies all locally elementary simplifications available in the current circuit. It is embedded inside the environment so that the agent never has to learn these simplifications for itself. After every agent action, the environment executes the CR routine before returning the next observation. The algorithm proceeds in two phases as shown in Figure~\ref{fig:CR}. In the \textit{commutation phase}, single-qubit gates are propagated toward one end of the circuit and CNOTs toward the other, using only commutation rules, until a commutation equilibrium is reached. This phase preserves the gate count but consolidates identical gates into contiguous clusters. In the \textit{reduction phase}, every available hard rule is applied greedily and exhaustively until no further reduction is possible. The CR algorithm guarantees that the circuit returned to the agent contains no reducible structure that is reachable by one simple commutation transformation, so the agent never observes a state with elementary simplifications still available.

\subsection{State Representation}
The state observed by the agent at each step is a structural representation of the current quantum circuit. It must encode all information needed to determine rule applicability, be compatible with the convolutional network architecture, and accommodate circuits of variable size. Each circuit is encoded as a three-dimensional tensor of shape $Q \times M \times C$~\cite{fosel2021quantum}, where $Q$ is the number of qubits, $M$ is the number of moments (parallel time slices), and $C$ is the total number of gate-type channels. Each single-qubit gate is represented by placing a $1$ at its corresponding $(q, m)$ coordinate within its designated channel. To encode the two-qubit CNOT gate accurately, two distinct channels are used to capture its spatial orientation: one channel represents a CNOT where the control qubit has a lower index than the target (control above target), and a second channel represents a CNOT where the control has a higher index than the target (control below target). When a CNOT is applied, a single $1$ is placed within the appropriate orientation channel at the spatial coordinate $q$ corresponding to the lowest qubit index involved in the gate (i.e., $q = \min(q_{\text{control}}, q_{\text{target}})$). Specifically, the channel count $C$ is computed as the number of single-qubit gate types in the chosen basis plus exactly two additional channels dedicated to the CNOT gate.

\subsection{Agent Architecture}

\begin{figure}[t]
    \centering
    \includegraphics[width=1.0\linewidth]{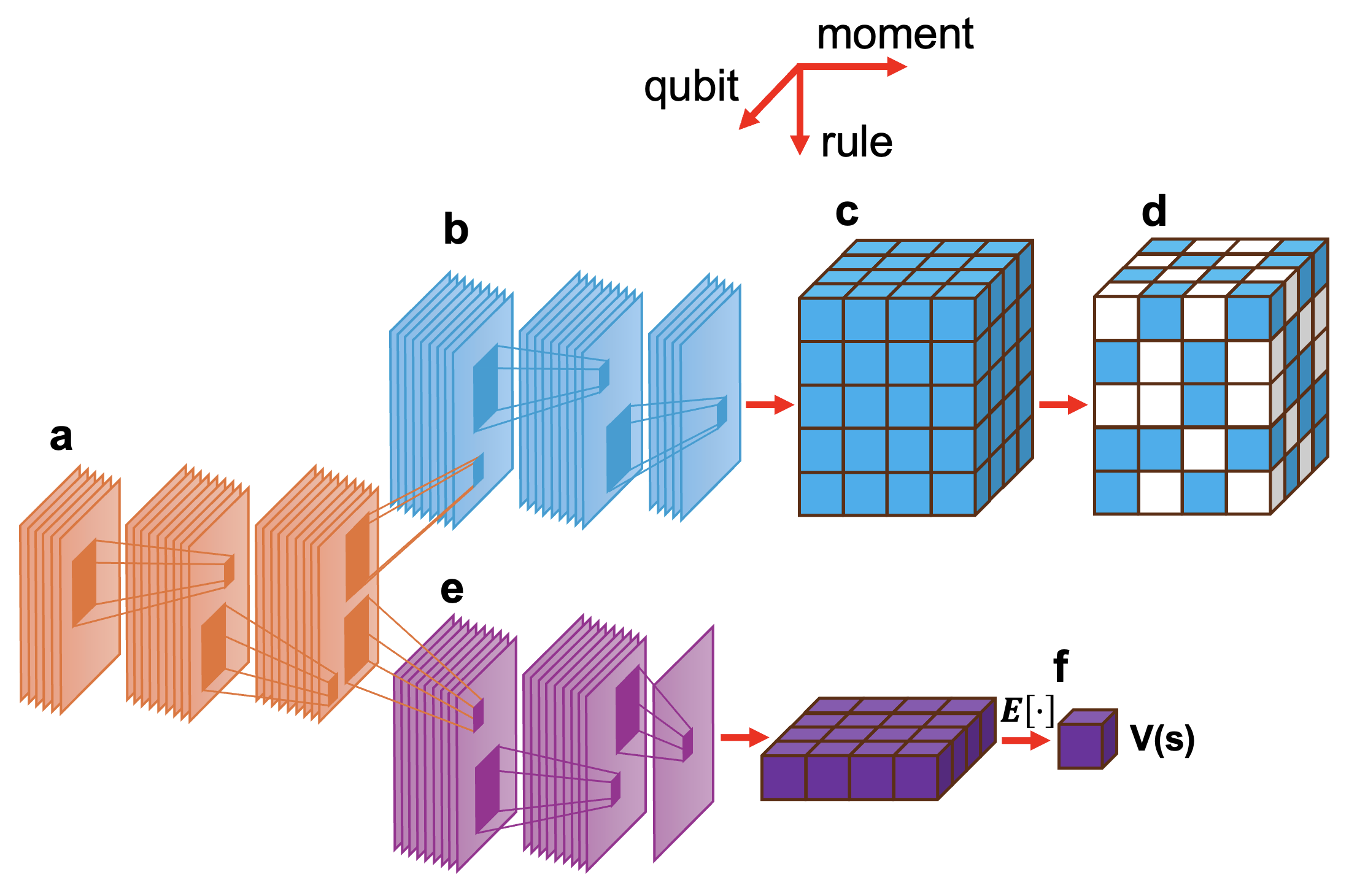}
    \caption {The Actor-Critic network architecture for quantum circuit optimization. (a) The shared convolutional backbone that extracts local spatio-temporal features. (b-c) The Actor (policy) branch, which generates a logit tensor of shape $Q \times M \times A$. (d) The application of the binary validity mask to ensure permissible action selection. (e-f) The Critic (value) branch, which aggregates features to produce a scalar state-value estimate $V(s)$.}
    \label{fig:agent}
\end{figure}

The agent is the central learning component of the system. It observes the current circuit and selects a transformation rule together with the location at which to apply it. The agent built on the Actor–Critic architecture used in this work follows~\cite{mnih2016asynchronous}. To accommodate circuits of arbitrary size, we adopt a fully convolutional architecture~\cite{long2015fully}. As shown in Figure \ref{fig:agent}, a shared backbone of two 2D convolutional layers extracts spatio-temporal features from the state tensor. The Actor head appends further convolutional layers to produce a logit tensor of shape $Q\times M\times N_R$, where $N_R=|\mathcal{R}|$ is the number of rules in the set of soft and decomposition rules $\mathcal{R}$. The output of the Actor head is interpreted as a joint distribution over the action triple $(q, m, k)$: the rule $k$ to apply, and the qubit and moment coordinates at which to apply the rule. The Critic head produces a single-channel value map and applies global average pooling over the spatial and temporal dimensions to yield a scalar state-value estimate $V(s)$.

At each step, the environment computes a binary validity mask of shape $Q \times M \times N_R$, with entry 1 if the corresponding rule is applicable at that location and 0 otherwise. The mask is applied to the Actor logits, so that probability mass is concentrated entirely on physically permissible actions.

\subsection{Reward Function}
The reward signal defines the agent's learning objective by quantifying the progress made toward circuit optimization at each time step. This progress is evaluated across a set of metrics $\mathcal{M}$, where a metric $\mu \in \mathcal{M}$ denotes any quantitative, non-negative structural property of a circuit $c$, expressed via a function $v_\mu(c)$, that characterizes its physical cost or complexity (e.g., total gate count, circuit depth, or two-qubit gate overhead). The reward at time step $t$ is defined as the weighted sum of improvements across $\mathcal{M}$
\begin{equation}
    r_t = \sum_{\mu \in \mathcal{M}} \alpha_\mu \left( v_\mu(t) - v_\mu(t+1) \right),
\end{equation}
where $v_\mu(t)$ represents the value of metric $\mu$ at time $t$, and $\alpha_\mu$ is a tunable coefficient that controls the relative importance of that metric. Within this generalized framework, the metric set $\mathcal{M} = \{\text{N}, \text{D}\}$ consists of the circuit's gate count $N$ and depth $D$, with corresponding values $v_\text{N}(t)$ and $v_\text{D}(t)$. Because reducing gate count inherently reduces circuit depth, the agent evaluated in this research prioritizes gate count over depth to minimize errors, setting $\alpha_\text{N} = 1$ and $\alpha_\text{D} = 0$. Since $v_\mu(t) - v_\mu(t+1)$ is positive whenever the metric decreases, the agent receives positive rewards upon metric reductions. Crucially, the state $v_\mu(t+1)$ is evaluated on the circuit returned by the CR algorithm, rather than the intermediate circuit produced immediately after the agent applies its chosen rule. Thus, the reward reflects the total reduction achieved by the combined effect of the agent's strategic action and the deterministic simplifications triggered by the CR algorithm.

\subsection{Training Data and Pipeline}

\begin{figure}[t]
    \centering
    \includegraphics[width=1.0\linewidth]{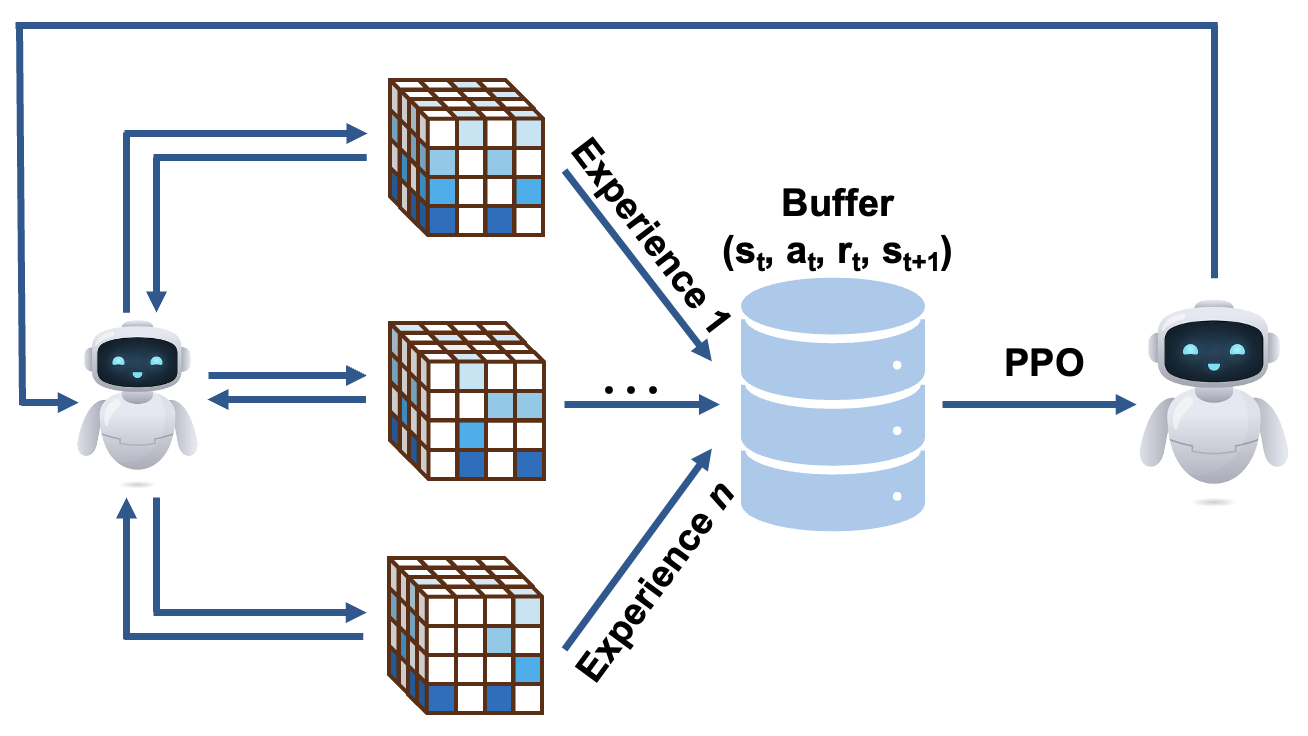}
    \caption {Schematic of the distributed experience collection and training pipeline. Multiple environments generate diverse circuit optimization trajectories, which are stored as transition tuples $(s_t, a_t, r_t, s_{t+1})$ in a centralized experience replay buffer. The buffered transitions are then sampled in shuffled mini-batches to train the policy with PPO.}
    \label{fig:training_pipeline}
\end{figure}

Effective training requires circuits that contain latent optimization opportunities, that is, simplifications not immediately accessible to greedy rule-based methods. We therefore generate training circuits using a five-stage protocol that systematically embeds such latent simplifications.
\begin{enumerate}
    \item Random Circuit Generation: An initial quantum circuit is generated by placing gates uniformly across qubits and moments.

    \item Initial Pruning: Deterministic hard rules are applied exhaustively to remove elementary redundancies.

    \item Expansion: To introduce new optimization possibilities, a CNOT-identity pair is inserted at a random location in the pruned circuit, exploiting the identity $\mathrm{CNOT} \cdot \mathrm{CNOT} = I$. The inserted pair is then commuted with adjacent single-qubit gates using the soft rules, which preserves the implemented unitary but spreads the two CNOTs across moments of the circuit. This increases both the gate count and the circuit depth, and creates a long-range pair of mutually cancellable CNOTs whose cancellation is no longer available under a single application of the hard rules.

    \item Re-pruning: The hard rules are applied again to remove any residual redundancies introduced by the expansion. The output is an expanded circuit that is locally irreducible but contains latent simplifications reachable only through strategic commutation.

    \item Iteration. The expansion and re-pruning cycle (Steps 3 and 4) is repeated for a fixed number of iterations.
\end{enumerate} 
The resulting circuits are functionally equivalent to their post-pruning baselines but contain reductions accessible only through strategic commutation, providing precisely the optimization problem the framework is designed to solve.

As shown in Figure \ref{fig:training_pipeline}, the agent gathers experience by repeatedly executing the standard reinforcement learning loop on multiple environments. These experience transitions, collected as tuples of $(s_t, a_t, r_t, s_{t+1})$, are recorded in a temporary rollout buffer. At the end of data collection, GAE is computed over the unbroken trajectories to respect temporal dependencies. For the training phase, the rollout buffer is randomly permuted and partitioned into mini-batches. By shuffling the experiences, the training data better approximates the independent and identically distributed (i.i.d.) assumption, leading to a more stable and effective learning process for the agent's neural networks. Hyperparameters used to train the agent are listed in the Appendix \ref{app:parameters}.

\section{Results}\label{sec_result}
\begin{figure*}[t]
    \centering
    \includegraphics[width=1.0\linewidth]{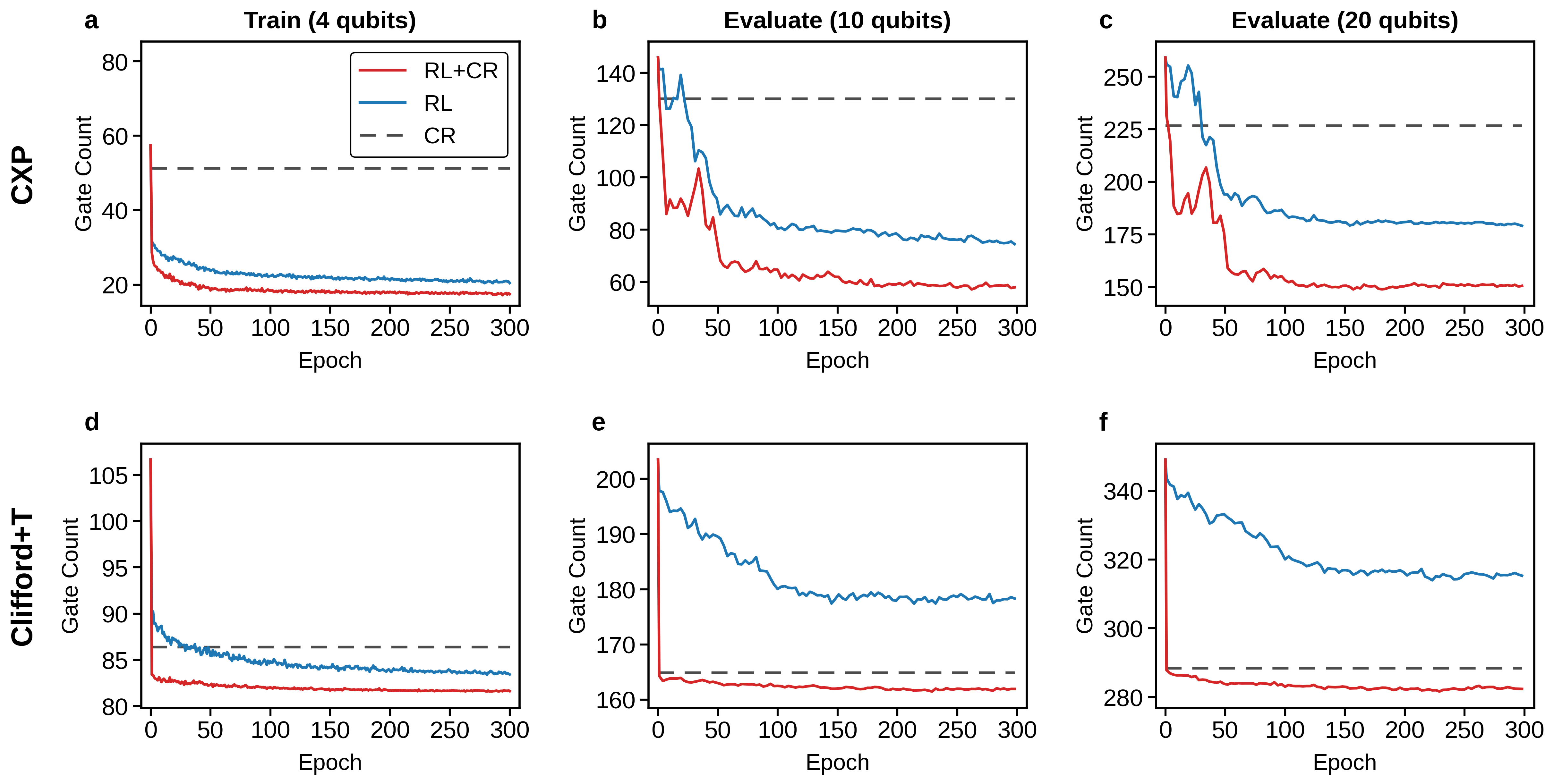}
    \caption {Training and evaluation learning curves for quantum circuit optimization. The top row (a–c) displays results for the CXP basis, while the bottom row (d–f) displays results for the Clifford+T basis. Column (a, d) represents the 4-qubit training circuits, while columns (b, e) and (c, f) demonstrate generalization to 10-qubit and 20-qubit circuits, respectively. In every panel, the RL-CR agent (blue) is compared against the standard RL agent (red) and the CR baseline (dash-dotted black). The vertical axis represents the average gate count, and the horizontal axis represents the training epochs.}
    \label{fig:train_val_performance}
\end{figure*}

We compare three methods across two gate sets, Clifford+T and CNOT-Pauli (CXP). The CR baseline applies the CR algorithm in isolation. The standard RL agent, hereafter RL, operates directly on raw circuits with no environment-side simplification. The proposed hybrid agent (RL+CR) operates on a CR-preprocessed environment, as described in Methods. For each gate set, separate agents are trained from scratch on 100 four-qubit circuits over 300 epochs. The trained policies are then tested, without any further training, on 50 ten-qubit circuits and 50 twenty-qubit circuits to assess how well a policy learned at 4 qubits transfers to circuits much larger (Table \ref{tab:circuit_stats}; learning curves in Figure \ref{fig:train_val_performance}).

\begin{table}[h]
\centering
\caption{Initial average gate counts of the training and evaluation circuit ensembles.}
\label{tab:circuit_stats}
\begin{tabular}{lcc}
\toprule
Circuit set & CXP & Clifford+T \\
\midrule
4-qubit (training)    & 57.36  & 106.65 \\
10-qubit (evaluation) & 145.86 & 203.48 \\
20-qubit (evaluation) & 259.06 & 349.08 \\
\bottomrule
\end{tabular}
\end{table}

RL+CR produces shorter circuits than both CR and RL in every condition tested. On 4-qubit CXP, RL+CR converges to an average of 16.93 gates, against 20.45 for RL and 51.24 for CR. On 20-qubit CXP, the unoptimized circuits average 259.06 gates; CR reduces this to 226.68 (12.5\% reduction), RL to 179.06 (30.9\% reduction), and RL+CR to 142.74 (44.9\% reduction) (Figure \ref{fig:train_val_performance}c). In absolute terms, RL+CR removed 116 gates per circuit on average, compared to 80 gates for the standard RL agent and 32 gates for CR alone (a $1.45 \times$ reduction of standard RL and a $3.6 \times$ that of CR). The advantage carries over to Clifford+T at 20 qubits (RL+CR: 282.36, CR: 288.34, RL: 315.30; Figure \ref{fig:train_val_performance}f), where the standard RL agent in fact fails to beat the CR baseline at either 10 or 20 qubits (Figure \ref{fig:train_val_performance}e, f).


The mechanism behind the RL+CR gap can be understood from the algorithm descriptions. The CR baseline applies a fixed deterministic procedure to the input circuit, namely the commutation phase followed by the hard-rule reduction phase, and stops once no further hard rule fires. Whatever cancellations the input circuit exposes under this procedure are found, and any cancellations that require a different rearrangement of the gates remain hidden. RL+CR adds one step in front of this procedure. At each iteration,
the agent applies a single strategic soft rule to the current circuit, rearranging it, and the CR algorithm then runs on the rearranged circuit. The agent's role is
to set up CR with a better starting point than the unmodified circuit would provide, that is, to expose cancellations that the input circuit does not. RL+CR and the standard RL agent have the same action space and the same network architecture. The only difference between them is that RL+CR does not have to learn the rules that CR already applies, since the CR algorithm performs them automatically after each agent action. The standard RL agent, by contrast, must rediscover these elementary reductions from scratch as part of its own training.


\begin{figure*}[t]
    \centering
    \includegraphics[width=0.9\linewidth]{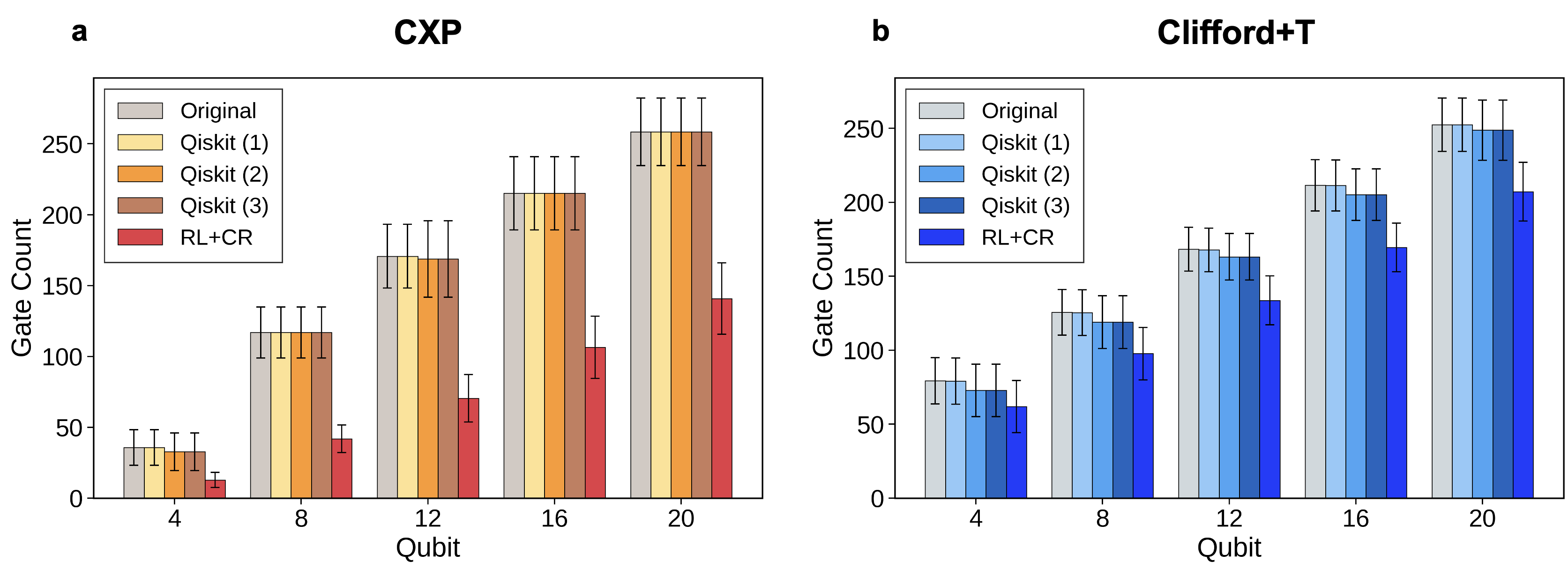}
    \caption {Optimization efficiency of RL+CR versus Qiskit's transpiler. Mean gate count after optimization across 50 evaluation circuits per system size, shown for the CXP (a) and Clifford+T (b) bases. Error bars indicate $\pm 1$ standard deviation. The original, unoptimized circuit (light grey) is shown for reference. We compare RL+CR against Qiskit's transpiler at all three standard optimization levels (\texttt{optimization\_level} = 1, 2, and 3 with \texttt{basis\_gates} restricted to the same gate set used in our framework). Cases in which Qiskit returned a circuit larger than the input are recorded as no reduction.}
    \label{fig:benchmark}
\end{figure*}

\begin{figure*}[t]
    \centering
    \includegraphics[width=1.\linewidth]{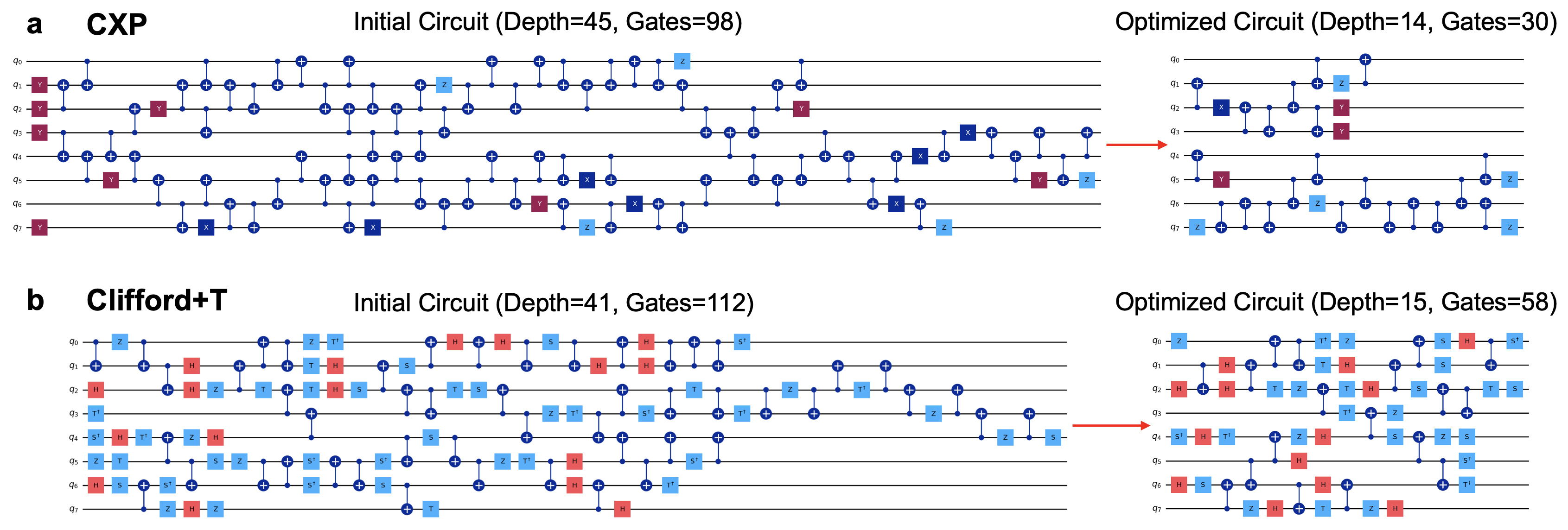}
    \caption {Reduction of quantum circuit depth and gate counts via the RL-CR approach. (a) Initial and optimized circuits evaluated on the Clifford+T basis. (b) Initial and optimized circuits evaluated on the CXP basis.}
    \label{fig:optimization}
\end{figure*}

To evaluate the optimization efficiency of our RL-based circuit reduction framework (RL+CR), we benchmarked it against Qiskit's transpiler at all three standard optimization levels (\texttt{optimization\_level} = 1, 2, and 3)~\cite{aleksandrowicz2019qiskit}. For each system size (4, 8, 12, 16, and 20 qubits) and each gate set, we evaluated 50 independent circuits, different from the training circuits, generated by the same protocol used during training. The original, unoptimized circuit gate counts are shown for reference. Across every system size and both bases, RL+CR produces strictly more compact circuits than Qiskit. The pattern is most pronounced in the CXP basis (Figure~\ref{fig:benchmark}a). We clearly demonstrate that Qiskit's transpiler achieves a modest 8.3\% reduction at 4 qubits but yields only marginal improvements at larger system sizes (no reduction at 8, 16, or 20 qubits, and 1.2\% at 12 qubits). RL+CR, in contrast, achieves 45.5--64.2\% reductions across the full range of benchmark circuits, with absolute savings over Qiskit growing monotonically from 19.9 fewer gates at 4 qubits to 117.7 fewer gates at 20 qubits. In the Clifford+T basis (Figure~\ref{fig:benchmark}b), Qiskit's reductions diminish with circuit size, from 8.2\% at 4 qubits to 1.5\% at 20 qubits. RL+CR maintains 18.0--22.2\% reductions throughout this range.

Figure~\ref{fig:optimization} illustrates an example of quantum circuit optimization, showing how the original circuit is transformed through the RL+CR algorithm. On the CXP basis, depth is compressed from 45 to 14 (69\% reduction) and gate count from 98 to 30 (69\% reduction). Comparable improvements are observed on the Clifford+T basis, where the algorithm reduces circuit depth from 41 to 15 and total gate count from 112 to 58, corresponding to reductions of 63\% and 48\%, respectively.
\section{Remarks}\label{sec_remarks}

We have introduced RL+CR, a framework for quantum circuit optimization that embeds a deterministic Commutation-and-Reduction algorithm directly into the training environment. By delegating elementary reductions to the CR routine, the agent no longer needs to rediscover simplifications that rule-based methods already resolve, and its learning capacity is concentrated on the non-trivial transformations where reinforcement learning genuinely adds value.

Empirical evaluation on two structurally distinct gate sets, the universal Clifford+T basis and the CXP basis, demonstrates three consistent advantages of this design. First, the trained policies produce strictly shorter circuits than both the CR baseline and the standard RL agent across all tested system sizes. Second, policies trained on 4-qubit circuits transfer effectively to circuits five times larger without any retraining. Benchmarking against Qiskit's highest-level transpiler pass further confirms that RL+CR achieves greater compression across both gate sets. Together, these findings establish that the proposed method is a robust and generalizable principle for quantum circuit optimization.

\textbf{Data and Code Availability}: The code implementing the RL+CR framework and the datasets generated and analyzed during the current study are available from the corresponding author upon reasonable request. 


\section*{Acknowledgement}

This work is supported by National Research Foundation of Korea (NRF) grant funded by the Korea government (MSIT) (RS-2024-00432113).

\newpage
\bibliography{reference}
\clearpage
\onecolumngrid
\appendix

\section{Network Architecture and Hyperparameters}
\label{app:parameters}

The specific layer configurations for each component are detailed in TABLE \ref{tab:full_arch}. The model accepts a 3D tensor representation of the circuit state with a shape denoted as (Batch, Channels, Qubits, Moments). The variables $Q$, $M$, and $A$ represent the number of qubits, moments, and available transformation rules, respectively.

\begin{table}[h!]
    \centering
    \small 
    \caption{Complete Network Architecture.}
    \label{tab:full_arch}
    \begin{tabular}{llc}
        \toprule
        \textbf{Layer} & \textbf{Details} & \textbf{Output Shape} \\
        \midrule
        
        \multicolumn{3}{l}{\textit{\textbf{Shared Feature Extractor}}} \\
        \midrule
        Input State & 3D Tensor Representation & ($B$, $C$, $Q$, $M$) \\
        Conv2D + Tanh & 256 filters, 3x3 kernel, padding 1 & ($B$, 256, $Q$, $M$) \\
        Conv2D + Tanh & 256 filters, 3x3 kernel, padding 1 & ($B$, 256, $Q$, $M$) \\
        \midrule

        \multicolumn{3}{l}{\textit{\textbf{Actor (Policy) Head}}} \\
        \midrule
        Input Features & From Shared Extractor & ($B$, 256, $Q$, $M$) \\
        Conv2D + Tanh & 256 filters, 3x3 kernel, padding 1 & ($B$, 256, $Q$, $M$) \\
        Conv2D (Logits) & $A$ filters, 3x3 kernel, padding 1 & ($B$, $A$, $Q$, $M$) \\
        \midrule

        \multicolumn{3}{l}{\textit{\textbf{Critic (Value) Head}}} \\
        \midrule
        Input Features & From Shared Extractor & ($B$, 256, $Q$, $M$) \\
        Conv2D + ReLU & 256 filters, 3x3 kernel, padding 1 & ($B$, 256, $Q$, $M$) \\
        Conv2D (Value Map) & 1 filter, 3x3 kernel, padding 1 & ($B$, 1, $Q$, $M$) \\
        Global Average Pool & Averages over Q and M dimensions & ($B$, 1) \\
        \bottomrule
    \end{tabular}
\end{table}

TABLE \ref{tab:hyperparams} lists the key hyperparameters selected to ensure stable and effective training with the PPO algorithm. The learning rate controls the size of gradient updates, while the discount factor $\gamma$ and GAE lambda $\lambda$ determine how the agent values future rewards and balances the bias-variance trade-off in its advantage estimates. The PPO-specific coefficients: clipping $\epsilon$, value loss $c_1$, and entropy $c_2$. We use the Adam optimizer to update the model weights.

\begin{table}[h!]
\centering
\caption{Training Hyperparameters.}
\label{tab:hyperparams}
\begin{tabular}{ll}
\toprule
\textbf{Hyperparameter} & \textbf{Value} \\
\midrule
Learning Rate & 5e-5 \\
Discount Factor ($\gamma$) & 0.99 \\
GAE Lambda ($\lambda$) & 0.95 \\
PPO Clipping ($\epsilon$) & 0.2 \\
Value Loss Coefficient ($c_1$) & 0.5 \\
Entropy Coefficient ($c_2$) & 0.01 \\
Gate Count Weight ($\alpha_N$) & 1 \\
Circuit Depth Weight ($\alpha_D$) & 0 \\
Batch Size & 256 \\
\bottomrule
\end{tabular}
\end{table}

\section{Transformation Rules}
\label{app:rules}

This appendix provides a comprehensive list of all quantum circuit transformations used in our reinforcement learning framework, together with the physical constraints under which they are applied. CNOT gates are restricted to adjacent qubits. The Pauli-Z gate is retained as an explicit element of the Clifford+T gate set, $Z \equiv S \cdot S$, allowing transformations to act on a $Z$ directly in a single step rather than on its two-$S$ equivalent.

The rules are divided into two categories based on their function. The first category, detailed in TABLE~\ref{tab:hard_rules_Clifford}, \ref{tab:hard_rules_Pauli}, consists of hard rules. These are deterministic transformations that guarantee a reduction in the circuit's cost. The second category, detailed in TABLE~\ref{tab:comm_exp_rules_Clifford}, \ref{tab:comm_exp_rules_Pauli}, comprises the soft and decomposition rules, which form the agent's learnable action space. Their purpose is to restructure the circuit to create new opportunities for subsequent optimizations. The transformations shown in the tables are representative. Each rule is depicted in one canonical orientation, but in practice, we apply all of its symmetric variants. Soft rules are bidirectional and may be applied in either direction.

The symbol $R_z$ in the tables below indicates the rotation gate around the Z axis for three special angles, namely $\pi/4$, $\pi/2$, and $\pi$. These angles correspond to the $T$, $S$, and Pauli-$Z$ gates, respectively. The symbol $R_z'$ denotes the Hermitian conjugate of $R_z$, i.e., the inverse rotation by the same angle, corresponding to $T^\dagger$, $S^\dagger$, and Pauli-Z, respectively.

\begin{figure}[htbp] 
\centering

\begin{minipage}[t]{0.45\textwidth}
    \centering
    \captionof{table}{Hard rules of Clifford+T basis}
    \label{tab:hard_rules_Clifford}
    \begin{tabular}{|c|c|c|}
        \toprule
        \textbf{ID} & \textbf{Before} & \textbf{After} \\
        \midrule
        H1 & \begin{quantikz} \gate{T} & \gate{T} \end{quantikz} & \begin{quantikz} \gate{S} \end{quantikz} \\ 
        \hline
        H2 & \begin{quantikz} \gate{T^\dagger} & \gate{T^\dagger} \end{quantikz} & \begin{quantikz} \gate{S^\dagger} \end{quantikz} \\
        \hline
        H3 & \begin{quantikz} \gate{S} & \gate{S} \end{quantikz} & \begin{quantikz} \gate{Z} \end{quantikz} \\
        \hline
        H4 & \begin{quantikz} \gate{S^\dagger} & \gate{S^\dagger} \end{quantikz} & \begin{quantikz} \gate{Z} \end{quantikz} \\
        \hline
        H5 & \begin{quantikz} \gate{S^\dagger} & \gate{Z} \end{quantikz} & \begin{quantikz} \gate{S} \end{quantikz} \\
        \hline
        H6 & \begin{quantikz} \gate{H} & \gate{H} \end{quantikz} & \begin{quantikz} \gate{I} \end{quantikz} \\
        \hline
        H7 & \begin{quantikz} \gate{S} & \gate{S^\dagger} \end{quantikz} & \begin{quantikz} \gate{I} \end{quantikz} \\
        \hline
        H8 & \begin{quantikz} \gate{T} & \gate{T^\dagger} \end{quantikz} & \begin{quantikz} \gate{I} \end{quantikz} \\
        \hline
        H9 & \begin{quantikz} \gate{Z} & \gate{Z} \end{quantikz} & \begin{quantikz} \gate{I} \end{quantikz} \\
        \hline
        H10 & \begin{quantikz} & \ctrl{1} & \ctrl{1} & \\ & \targ{} & \targ{} & \end{quantikz} & \begin{quantikz} \gate{I} \\ \gate{I} \end{quantikz} \\
        \hline
        H11 & \begin{quantikz} \gate{H} & \ctrl{1} & \gate{H} \\ \gate{H} & \targ{} & \gate{H} \end{quantikz} & \begin{quantikz} & \targ{} & \\ & \ctrl{-1} & \end{quantikz} \\
        \hline
        H12 & \begin{quantikz} \gate{S} & \ctrl{1} &  &  \ctrl{1} &\\ \gate{S} & \targ{} & \gate{S^\dagger} & \targ{} & \end{quantikz} & \begin{quantikz}  & \ctrl{1} &  \\ \gate{H} & \targ{} & \gate{H} \end{quantikz} \\
        \hline
        H13 & \begin{quantikz} & \ctrl{1} & \gate{Z} \\ & \targ{} & \gate{Z} \end{quantikz} & \begin{quantikz} & \ctrl{1} & \\ \gate{Z} & \targ{} &\end{quantikz} \\
        \hline
        H14 & \begin{quantikz} \gate{H} & \gate{S} & \targ{} & \gate{S^\dagger} & \gate{H} \\  &  & \ctrl{-1} &  &  \end{quantikz} & \begin{quantikz} \gate{S^\dagger} & \targ{} & \gate{S} \\  & \ctrl{-1} &  \end{quantikz} \\
        \hline
        H15 & \begin{quantikz} \gate{H} & \gate{S^\dagger} & \targ{} & \gate{S} & \gate{H} \\  &  & \ctrl{-1} &  &  \end{quantikz} & \begin{quantikz} \gate{S} & \targ{} & \gate{S^\dagger} \\ & \ctrl{-1} & \end{quantikz} \\
        \bottomrule
    \end{tabular}
\end{minipage}
\hfill 
\begin{minipage}[t]{0.50\textwidth}
    \centering
    \captionof{table}{Soft and Decomposition Rules of Clifford+T basis}
    \label{tab:comm_exp_rules_Clifford}
    \begin{tabular}{|c|c|c|}
        \toprule
        \textbf{ID} & \textbf{Before} & \textbf{After} \\
        \midrule
        S1 & \begin{quantikz} \gate{T} & \gate{S} \end{quantikz} & \begin{quantikz} \gate{S} & \gate{T} \end{quantikz} \\
        \hline
        S2 & \begin{quantikz} \gate{T} & \gate{S^\dagger} \end{quantikz} & \begin{quantikz} \gate{S^\dagger} & \gate{T} \end{quantikz} \\
        \hline
        S3 & \begin{quantikz} \gate{T^\dagger} & \gate{S} \end{quantikz} & \begin{quantikz} \gate{S} & \gate{T^\dagger} \end{quantikz} \\
        \hline
        S4 & \begin{quantikz} \gate{T^\dagger} & \gate{S^\dagger} \end{quantikz} & \begin{quantikz} \gate{S^\dagger} & \gate{T^\dagger} \end{quantikz} \\
        \hline
        S5 & \begin{quantikz} \gate{T} & \gate{Z} \end{quantikz} & \begin{quantikz} \gate{Z} & \gate{T} \end{quantikz} \\
        \hline
        S6 & \begin{quantikz} \gate{T^\dagger} & \gate{Z} \end{quantikz} & \begin{quantikz} \gate{Z} & \gate{T^\dagger} \end{quantikz} \\
        \hline
        S7 & \begin{quantikz} \gate{S} & \gate{Z} \end{quantikz} & \begin{quantikz} \gate{Z} & \gate{S} \end{quantikz} \\
        \hline
        S8 & \begin{quantikz} \gate{Rz} & \ctrl{1} &\\  & \targ{} &\end{quantikz} & \begin{quantikz} & \ctrl{1} & \gate{Rz}\\ & \targ{} & \end{quantikz} \\
        \hline
        S9 & \begin{quantikz} & \targ{} &  &\\ & \ctrl{-1} & \ctrl{1} & \\ &  & \targ{} &\end{quantikz} & \begin{quantikz} &  & \targ{} &\\ & \ctrl{1} & \ctrl{-1} & \\ & \targ{} &  & \end{quantikz} \\
        \hline
        S10 & \begin{quantikz} & \ctrl{1} &  &\\ & \targ{} & \targ{} & \\ &  & \ctrl{-1} &\end{quantikz} & \begin{quantikz} &  & \ctrl{1} &\\ & \targ{} & \targ{} & \\ & \ctrl{-1} &  & \end{quantikz} \\
        \hline
        S11 & \begin{quantikz} & \ctrl{1} &  &  \ctrl{1} &\\ & \targ{} & \gate{Rz} & \targ{} & \end{quantikz} & \begin{quantikz} & \targ{} & \gate{Rz} & \targ{} &\\ & \ctrl{-1} &  & \ctrl{-1} & \end{quantikz} \\
        \hline
        S12 & \begin{quantikz} &  & \ctrl{1} &\\ \gate{Rz} & \gate{H} & \targ{} & \gate{H} \end{quantikz} & \begin{quantikz} & \ctrl{1} &  & \\ \gate{H} & \targ{} & \gate{H} & \gate{Rz} \end{quantikz} \\
        \hline
        S13 & \begin{quantikz} & \ctrl{1} &  &  \ctrl{1} &\\ \gate{Rz} & \targ{} & \gate{R'z} & \targ{} & \end{quantikz} & \begin{quantikz} & \ctrl{1} &  &  \ctrl{1} &\\ \gate{R'z} & \targ{} & \gate{Rz} & \targ{} & \end{quantikz} \\
        \hline
        S14 & \begin{quantikz} & \ctrl{1} &  &  &\\ & \targ{} & \gate{H} & \ctrl{1} & \gate{H}\\ &   &  & \targ{} & \end{quantikz} & \begin{quantikz} &  &  & \ctrl{1} &\\ \gate{H} & \ctrl{1} & \gate{H} & \targ{} & \\ & \targ{} &  &  & \end{quantikz} \\
        \hline
        D1 & \begin{quantikz} & \targ{} &\\ & \ctrl{-1} &\end{quantikz} & \begin{quantikz} \gate{H} & \ctrl{1} & \gate{H}\\ \gate{H} & \targ{} & \gate{H} \end{quantikz} \\
        \bottomrule
    \end{tabular}
\end{minipage}
\end{figure}

\begin{figure}[htbp] 
\centering

\begin{minipage}[t]{0.43\textwidth}
    \centering
    \captionof{table}{Hard rules of CXP basis}
    \label{tab:hard_rules_Pauli}
    \begin{tabular}{|c|c|c|}
        \toprule
        \textbf{ID} & \textbf{Before} & \textbf{After} \\
        \midrule
        H1 & \begin{quantikz} \gate{X} & \gate{X} \end{quantikz} & \begin{quantikz} \gate{I} \end{quantikz} \\
        \hline
        H2 & \begin{quantikz} \gate{Y} & \gate{Y} \end{quantikz} & \begin{quantikz} \gate{I} \end{quantikz} \\
        \hline
        H3 & \begin{quantikz} \gate{Z} & \gate{Z} \end{quantikz} & \begin{quantikz} \gate{I} \end{quantikz} \\
        \hline
        H4 & \begin{quantikz} \gate{X} & \gate{Y} \end{quantikz} & \begin{quantikz} \gate{iZ} \end{quantikz} \\
        \hline
        H5 & \begin{quantikz} \gate{Y} & \gate{X} \end{quantikz} & \begin{quantikz} \gate{-iZ} \end{quantikz} \\
        \hline
        H6 & \begin{quantikz} \gate{Z} & \gate{X} \end{quantikz} & \begin{quantikz} \gate{iY} \end{quantikz} \\
        \hline
        H7 & \begin{quantikz} \gate{X} & \gate{Z} \end{quantikz} & \begin{quantikz} \gate{-iY} \end{quantikz} \\
        \hline
        H8 & \begin{quantikz} \gate{Y} & \gate{Z} \end{quantikz} & \begin{quantikz} \gate{iX} \end{quantikz} \\
        \hline
        H9 & \begin{quantikz} \gate{Z} & \gate{Y} \end{quantikz} & \begin{quantikz} \gate{-iX} \end{quantikz} \\
        \hline
        H10 & \begin{quantikz} & \ctrl{1} & \ctrl{1} & \\ & \targ{} & \targ{} & \end{quantikz} & \begin{quantikz} \gate{I} \\ \gate{I} \end{quantikz} \\
        \hline
        H11 & \begin{quantikz} & \ctrl{1} & \gate{X} \\ & \targ{} & \gate{X} \end{quantikz} & \begin{quantikz} \gate{X} & \ctrl{1} & \\ & \targ{} & \end{quantikz} \\
        \hline
        H12 & \begin{quantikz} & \ctrl{1} & \gate{Z} \\ & \targ{} & \gate{Z} \end{quantikz} & \begin{quantikz} & \ctrl{1} & \\ \gate{Z} & \targ{} &\end{quantikz} \\
        \hline
        H13 & \begin{quantikz} & \ctrl{1} & \gate{Y} \\ & \targ{} & \gate{X} \end{quantikz} & \begin{quantikz} \gate{Y} & \ctrl{1} & \\ & \targ{} &\end{quantikz} \\
        \hline
        H14 & \begin{quantikz} & \ctrl{1} & \gate{Z} \\ & \targ{} & \gate{Y} \end{quantikz} & \begin{quantikz} & \ctrl{1} & \\ \gate{Y} & \targ{} &\end{quantikz} \\
        \bottomrule
    \end{tabular}
\end{minipage}
\hfill 
\begin{minipage}[t]{0.55\textwidth}
    \centering
    \captionof{table}{Soft and Decomposition Rules of CXP basis}
    \label{tab:comm_exp_rules_Pauli}
    \begin{tabular}{|c|c|c|}
        \toprule
        \textbf{ID} & \textbf{Before} & \textbf{After} \\
        \midrule
        S1 & \begin{quantikz} \gate{Z} & \ctrl{1} &\\  & \targ{} &\end{quantikz} & \begin{quantikz} & \ctrl{1} & \gate{Z}\\ & \targ{} & \end{quantikz} \\
        \hline
        S2 & \begin{quantikz} & \ctrl{1} &\\ \gate{X} & \targ{} &\end{quantikz} & \begin{quantikz} & \ctrl{1} & \\ & \targ{} & \gate{X}\end{quantikz} \\
        \hline
        S3 & \begin{quantikz} & \targ{} &  &\\ & \ctrl{-1} & \ctrl{1} & \\ &  & \targ{} &\end{quantikz} & \begin{quantikz} &  & \targ{} &\\ & \ctrl{1} & \ctrl{-1} & \\ & \targ{} &  & \end{quantikz} \\
        \hline
        S4 & \begin{quantikz} & \ctrl{1} &  &\\ & \targ{} & \targ{} & \\ &  & \ctrl{-1} &\end{quantikz} & \begin{quantikz} &  & \ctrl{1} &\\ & \targ{} & \targ{} & \\ & \ctrl{-1} &  & \end{quantikz} \\
        \hline
        D1 & \begin{quantikz} \gate{X} & \ctrl{1} & \\ & \targ{} & \end{quantikz} & \begin{quantikz} & \ctrl{1} & \gate{X} \\ & \targ{} & \gate{X} \end{quantikz} \\
        \hline
        D2 & \begin{quantikz} \gate{Y} & \ctrl{1} & \\ & \targ{} & \end{quantikz} & \begin{quantikz} & \ctrl{1} & \gate{Y} \\ & \targ{} & \gate{X} \end{quantikz} \\
        \hline
        D3 & \begin{quantikz} & \ctrl{1} & \\ \gate{Y} & \targ{} & \end{quantikz} & \begin{quantikz} & \ctrl{1} & \gate{Z} \\ & \targ{} & \gate{Y} \end{quantikz} \\
        \hline
        D4 & \begin{quantikz} & \ctrl{1} & \\ \gate{Z} & \targ{} & \end{quantikz} & \begin{quantikz} & \ctrl{1} & \gate{Z} \\ & \targ{} & \gate{Z} \end{quantikz} \\
        \bottomrule
    \end{tabular}
\end{minipage}
\end{figure}

\end{document}